\documentclass[fleqn,12pt,twoside]{article}
\usepackage{espcrc1}

\usepackage{graphicx}
\newcommand{\AmS}{{\protect\the\textfont2
  A\kern-.1667em\lower.5ex\hbox{M}\kern-.125emS}}

\title{Modified Fragmentation Function and Jet Quenching at RHIC}

\author{Xin-Nian Wang\address{Nuclear Science Division, Mailstop 70-319, 
         Lawrence Berkeley Laboratory, Berkeley, California 94720,USA }}

\begin{document}
%%LBNL-51282

% typeset front matter
\maketitle

\begin{abstract}
Medium modification of jet fragmentation functions and parton energy loss 
in cold and hot matter are reviewed. The predicted nuclear modification
of the jet fragmentation function agrees well with the recent HERMES
data with a resultant energy loss $dE/dx\approx 0.5$ GeV/fm. From the
the recent PHENIX data of high $p_T$ $\pi_0$ spectra in central
$Au+Au$ collisions at $\sqrt{s}=130$ GeV, one extracts an energy
loss for a 10 GeV parton that is equivalent to $dE/dx=7.3$ GeV/fm in 
a static medium with the same gluon density as in the initial 
stage of the collision at
$\tau_0=0.2$ fm/$c$. Constraints on jet quenching by the central
rapidity density of charged hadrons is also discussed.
\end{abstract}

\section{INTRODUCTION}

It is with a great respect that I present this talk on this
special occasion of celebrating Helmut Satz's scientific career and his
unwavering role in high-energy heavy-ion physics. I first met Helmut
when I was still a graduate student at my first Quark Matter meeting
in 1988. His name was already synonymous to quark-gluon plasma and
$J/\Psi$ suppression for me at that time. Because of our common interests
in hard processes in heavy-ion collisions, we started working together
in 1994 to coordinate a working group called Hard Probes Collaboration.
Since then this working group has become a forum where we meet regularly
together with other colleagues to discuss problems related to hard
processes in high-energy heavy-ion collisions. Helmut has been an
inspiration for all of us in the working group and in the community
of high-energy nuclear physics.

The subject of jet quenching and parton energy loss is one
aspect of hard processes in heavy-ion collisions. However, with the
advent of the RHIC facility and experiments, jets of high-energy partons 
will become an important and useful tool to the study of the properties
of dense matter formed in high-energy heavy-ion collisions. Because
large $p_T$ partons are produced very early in heavy-ion collisions
and their production rates can be calibrated in $pp$ and $pA$ collisions
at the same energy, they are ideal probes of the dense matter that
is formed in the same reaction. What probes the dense medium is the
scattering induced energy loss suffered by an energetic parton as it 
propagates through the matter. The parton energy loss is directly
related to the parton density of the medium.

Theoretical studies of the parton energy loss in hot medium date back to 
the first attempt by Bjorken \cite{bj} to calculate elastic energy
loss of a parton via elastic scattering in the hot medium. A simple
estimate can be given by the thermal averaged energy transfer
$\nu_{\rm el}\approx q_\perp^2/2\omega$ of the jet parton to a thermal
parton with energy $\omega$, $q_\perp$ being the transverse momentum
transfer of the elastic scattering. The resultant elastic energy 
loss \cite{wangrep}
\begin{equation}
\frac{dE_{\rm el}}{dx}=C_2\frac{3\pi\alpha_{\rm s}^2}{2}T^2
\ln\left(\frac{3ET}{2\mu^2}\right)
\end{equation}
is sensitive to the temperature of the thermal medium but is in general 
small compared to radiative energy loss. Here, $\mu$ is the Debye screening
mass and $C_2$ is the Casimir of the propagating parton in its fundamental 
presentation. The elastic energy loss can also be calculated within finite
temperature QCD \cite{thoma} with a similar result, but with a more careful
and consistent treatment of screening effect.

Though there had been estimates of
the radiative parton energy loss using the uncertainty principle \cite{bh},
a first theoretical study of QCD radiative parton energy loss incorporating
Landau-Pomeranchuk-Migdal interference effect \cite{lpm} is by Gyulassy and
myself \cite{gw} where multiple parton scattering is modeled by a screened
Coulomb potential model. Baier {\it et al.} (BDMPS) \cite{bdmps} later 
considered the effect of gluon rescattering which turned out to be very 
important for gluon radiation induced by multiple scattering in a dense
medium. These two studies have ushered in many recent works on the subject,
including a path integral approach to the problem \cite{zakharov} and opacity 
expansion framework \cite{glv,wied} which is more suitable for multiple
parton scattering in a thin plasma. The radiative parton energy loss to the
leading order of the opacity $\bar{n}=L/\lambda$ in the thin plasma of size
$L$ is estimated as \cite{glv,ww}
\begin{equation}
\frac{dE_{\rm rad}}{dx}\approx C_2 \frac{\alpha_{\rm s} \mu^2}{4}
\frac{L}{\lambda} \ln\left(\frac{2E}{\mu^2 L}\right),
\end{equation}
where $\lambda$ is the gluon's mean-free-path in the medium. The unique
$L$ dependence of the parton energy loss is a consequence of the non-Abelian
LMP interference effect in a QCD medium. It is also shown in a recent
study \cite{ww} that thermal absorption and stimulated emission in a thermal
environment can be neglected for high energy partons ($E\gg \mu$) while
they are important for intermediate energy partons.

Using this latest result one can estimate the total energy loss for 
a parton with initial energy $E=40$ GeV to be about $\Delta E\approx 10$ GeV
after it propagates a distance of $L=6$ fm in a medium with $\mu=0.5$ GeV
and $\lambda=1$ fm. For an expanding system, the total energy loss
is reduced by a factor of $2\tau_0/L$ from the static 
value \cite{baier98,gvw}. Assuming that most of
this energy loss is carried by gluons outside the jet cone \cite{baier99},
measuring the energy loss would require the experimental resolution
$\delta E$ to be much smaller than the total energy loss $\Delta E$.
With the measured total multiplicity density $dN/d\eta \approx 900$
\cite{phobos} and energy density $dE_T/d\eta\approx 500$ GeV \cite{phet}
in central $Au+Au$ collisions at $\sqrt{s}=130$ GeV, one can estimate that
the average total background energy within the jet cone ($\delta\eta=1$
and $\delta\phi=1$) is about $\Sigma E_T\approx 80$ GeV with
a fluctuation of $\delta E_T\approx 10$ GeV. It is therefore very
difficult, if not impossible, to determine the energy of a jet on
a event-by-event base \cite{Wang:1990bk}. Since high $p_T$ hadrons
in hadron and nuclear collisions come from fragmentation of high $E_T$
jets, energy loss naturally leads to suppression of high $p_T$ hadron
spectra. Miklos Gyulassy and I then proposed \cite{wg92} that one has 
to reply on measuring the suppression of high $p_T$ hadrons to study 
parton energy loss in heavy-ion collisions. Since inclusive hadron
spectra is a convolution of jet production cross section and the
jet fragmentation function in pQCD, the suppression of inclusive high $p_T$
hadron spectra is a direct consequence of the medium modification of the
jet fragmentation function induced by parton energy loss.
Assuming that jet fragmentation function is the same for the final
leading parton with a reduced energy, the modified fragmentation 
function can be assumed as \cite{wh}
\begin{equation}
\widetilde{D}(z)\approx \frac{1}{1-\Delta E/E} 
D\left(\frac{z}{1-\Delta E/E}\right). \label{dbar0}
\end{equation}
Therefore, in this effective model the measured modification of fragmentation 
function can be directly related to the parton energy loss.

\section{MODIFIED FRAGMENTATION FUNCTIONS}

Since a jet parton is always produced via a hard process involving a
large momentum scale, it should also have final state radiation with
and without rescattering leading to the DGLAP evolution equation of
fragmentation functions. Such final state radiation effectively
acts as a self-quenching mechanism softening the leading parton
momentum distribution. This process is quite similar to the induced 
gluon radiation and the two should have strong interference
effect \cite{glv,wied}. It is therefore natural to study jet quenching
and modified fragmentation function in the framework of modified
DGLAP evolution equations in a medium \cite{wgdis}.

The simplest case for jet quenching is deeply inelastic scattering
of an electron off a nucleus target where the virtual photon knocks
one quark out of a nucleon inside a nucleus. The quark then will have
to propagate through the rest of the nucleus and possibly scatter
again with other nucleons with induced gluon radiation. The
induced gluon radiation reduces the quark's energy before it fragments 
into hadrons with a modified fragmentation function. One can
study the nuclear modification of the fragmentation function by
comparing it with the same measurement in DIS with a nucleon target.

\begin{figure}[htb]
\begin{minipage}[t]{80mm}
\includegraphics[scale=0.8]{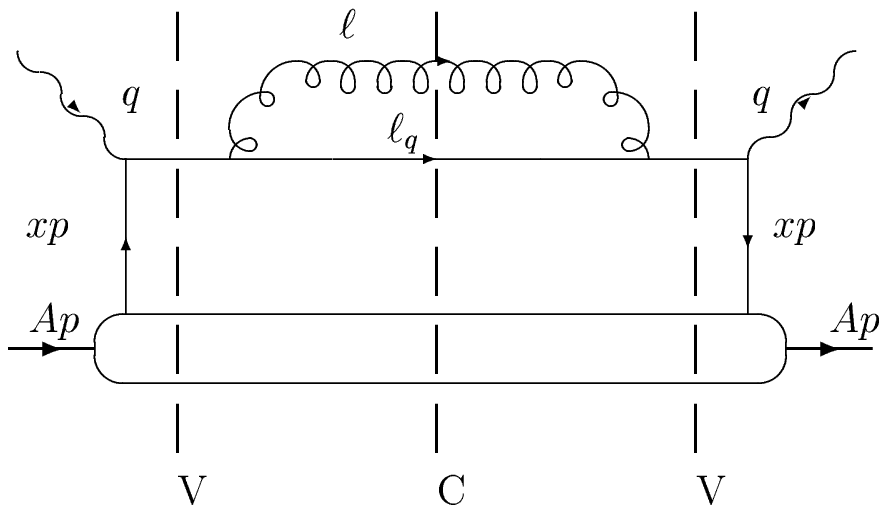}
\caption{Diagram of final state radiation correction to the jet
fragmentation function with central and virtual cuts.
}
\label{fig1}
\end{minipage}
\hspace{\fill}
\begin{minipage}[t]{75mm}
\includegraphics[scale=0.8]{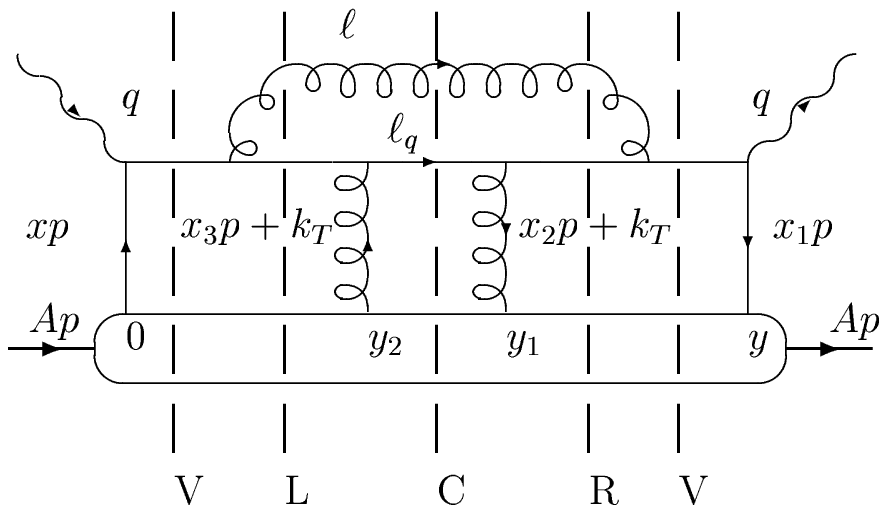}
\caption{One example diagram of gluon radiation associated with double
scattering with different cuts (central, left, right and virtual).
}
\label{fig2}
\end{minipage}
\end{figure}

We work in an infinite momentum frame, where the photon carries momentum
$q=[-x_Bp^+,q^-,\vec{0}_\perp]$ and the momentum of the target
per nucleon is $p=[p^+,0,\vec{0}_\perp]$ with the Bjorken variable
defined as $x_B=Q^2/2q^-p^+$. The differential semi-inclusive
hadronic tensor in a collinear factorization approximation to the
leading twist can be written as
\begin{equation}
\frac{dW_{\mu\nu}^{S}}{dz_h}
= \sum_q \int dx f_q^A(x,\mu^2_I) H_{\mu\nu}(x,p,q) 
D_{q\rightarrow h}(z_h,\mu^2)\ ,
\label{eq:w-s}
\end{equation}
where $f_q^A(x,\mu^2_I)$ is the quark distribution of the nucleus,
$H_{\mu\nu}(x,p,q)$ is the hard part of $\gamma^*+q$ scattering
and $D_{q\rightarrow h}(z_h,\mu^2)$ is the quark fragmentation
function in vacuum as measured in the $e^+e^-$ annihilation process.
The scale $\mu^2$ dependence of the fragmentation function comes
from the radiative correction as shown in Fig.~\ref{fig1}. Taking
into account of both the radiative (central cut diagram) and
the virtual correction (virtual cut diagram), the renormalized
fragmentation function is defined as,
\begin{eqnarray}
D_{q\rightarrow h}(z_h,\mu^2)\equiv D_{q\rightarrow h}(z_h) &+&
\int_0^{\mu^2} \frac{d\ell_T^2}{\ell_T^2} 
\frac{\alpha_s}{2\pi} \int_{z_h}^1 \frac{dz}{z} \nonumber \\
&\times&
\left[\gamma_{q\rightarrow qg}(z) D_{q\rightarrow h}(z_h/z)
+\gamma_{q\rightarrow gq}(z) D_{g\rightarrow h}(z_h/z)\right],
\label{eq:s-sum}
\end{eqnarray}
which satisfies the DGLAP evolution equation,
\begin{equation}
  \frac{\partial D_{q\rightarrow h}(z_h,\mu^2)}{\partial \ln \mu^2}=
  \frac{\alpha_s}{2\pi} \int^1_{z_h} \frac{dz}{z} 
\left[ \gamma_{q\rightarrow qg}(z)
D_{q\rightarrow h}(z_h/z,\mu^2) + \gamma_{q\rightarrow gq}(z) 
D_{g\rightarrow h}(z_h/z,\mu^2)\right].  \label{eq:ap1}
\end{equation}
Here $\gamma_{q\rightarrow qg}(z)=\gamma_{q\rightarrow gq}(1-z)
=C_F[(1+z^2)/(1-z)_+ +(3/2)\delta(1-z)]$ is the splitting function.

In a nucleus target, the outgoing quark can scatter again with another
parton from the nucleus. The additional scattering
may induce additional gluon radiation and cause the leading quark to 
lose energy. Such induced gluon radiation will effectively give 
rise to additional terms in the evolution equation leading to modification 
of the fragmentation functions in a medium. Contributions from 
multiple parton scattering are always non-leading twist. However
we will consider only those that are enhanced by the nuclear size $A^{1/3}$.

We work in the LQS framework \cite{lqs} in which the twist-four 
contributions can be expressed as the convolution of partonic hard parts
and four-parton matrix elements. At the lowest order 
(processes without gluon radiation) in this framework, rescattering 
with collinear gluons gives the eikonal contribution to the
gauge-invariant leading-twist and lowest-order result, 
assuming collinear factorization of the quark fragmentation function.
The radiative correction to double scattering processes with another 
gluon from the nucleus, shown in Fig.~\ref{fig2} as an example, generally
involves matrix elements of four parton fields and the cross section
of four parton scattering. In the LQS framework, one makes collinear
expansion of the four-parton scattering cross section in terms of the
transverse momentum of the scattering gluons. The first term in the 
collinear expansion corresponds to rescattering with collinear gluons.
Similar to the leading order case, it contributes only to the
eikonal correction of the radiative processes at the leading twist
as a consequence of the cancellation between different cut diagrams
(central, right and left cut). The leading twist-four contribution
then comes from the second derivative of the four-parton scattering
cross section in the central-cut diagrams. With the quadratic 
term $k_\perp^\alpha k_\perp^\beta$, this contribution depends on 
the gluon field strength, because
$k_\perp^{\alpha}A^+k_\perp^{\beta}A^+ \rightarrow 
F^{\alpha+}F^{\beta+}$ by partial integrations.

Considering all possible diagrams in addition to the example
shown in Fig.~\ref{fig2}, including the virtual corrections (virtual cut), 
one get the leading twist-four contribution to the semi-inclusive cross 
section from double scattering,
\begin{eqnarray}
\frac{W_{\mu\nu}^{D}}{dz_h}
&=&\sum_q \,\int dx H^{(0)}_{\mu\nu}(xp,q)
\frac{2\pi\alpha_s}{N_c}\int \frac{d\ell_T^2}{\ell_T^4}  
\int_{z_h}^1\frac{dz}{z} \frac{\alpha_s}{2\pi} C_A  \nonumber \\
&\times& \left\{ D_{q\rightarrow h}(z_h/z) \left[
\frac{1+z^2}{(1-z)_+}T^A_{qg}(x,x_L)
+\delta(z-1)\Delta T^A_{qg}(x,\ell_T^2) \right]\right. \nonumber \\
&+& \left. D_{g\rightarrow h}(z_h/z) \left[
\frac{1+(1-z)^2}{z_+}T^A_{qg}(x,x_L)+\delta(z)
\Delta T^A_{qg}(x,\ell_T^2) \right] \right\}, \label{eq:WD-fq2}
\end{eqnarray}
where 
\begin{eqnarray}
T^A_{qg}(x,x_L)&=& \int \frac{dy^{-}}{2\pi}\, dy_1^-dy_2^-
e^{i(x+x_L)p^+y^-+ix_Tp^-(y_1^--y_2^-)}(1-e^{-ix_Lp^+y_2^-})
(1-e^{-ix_Lp^+(y^--y_1^-)}) \nonumber \\
&\frac{1}{2}&\langle A | \bar{\psi}_q(0)\,
\gamma^+\, F_{\sigma}^{\ +}(y_{2}^{-})\, F^{+\sigma}(y_1^{-})\,\psi_q(y^{-})
| A\rangle \theta(-y_2^-)\theta(y^--y_1^-)
\label{eq:qgmatrix}
\end{eqnarray}
is the quark-gluon correlation function which essentially 
contains four independent twist-four parton matrix elements in a nucleus.
Here $x_L=\ell_\perp^2/2p^+q^-z(1-z)$ and 
$x_T=\langle k_T^2\rangle/2p^+q^-z(1-z)$.
In the central-cut diagrams (Fig.~\ref{fig2} for example) where the
leading higher-twist result comes from, there are typically four contributions
from each cut diagram. The radiated gluon can either come as the final
state radiation of the $\gamma$-quark scattering or as the initial
state radiation of the secondary quark-gluon scattering. The
amplitudes of these initial and final state radiation have opposite signs
with also a phase difference $x_Lp^+y^-$. 
The sum of these two radiation processes and their interferences
gives rise to the dipole-like factor 
$(1-e^{-ix_Lp^+y_2^-})(1-e^{-ix_Lp^+(y^--y_1^-)})$ in the effective
two-parton correlation function that enters into the double-scattering
cross section with induced gluon radiation. This is exactly the LPM
effect of bremsstrahlung in medium.

Summing up the single and double scattering contributions to the
semi-inclusive process, one can define the effective modified
fragmentation function as
\begin{equation}
\frac{dW_{\mu\nu}}{dz_h}=\frac{dW_{\mu\nu}^S}{dz_h}+\frac{dW_{\mu\nu}^D}{dz_h}
=\sum_q \int dx f_q^A(x,\mu_I^2) 
H^{(0)}_{\mu\nu}(x,p,q)
\widetilde{D}_{q\rightarrow h}(z_h,\mu^2), \label{eq:Wtot}
\end{equation}
\begin{eqnarray}
\widetilde{D}_{q\rightarrow h}(z_h,\mu^2)&\equiv& 
D_{q\rightarrow h}(z_h,\mu^2) 
+\Delta D_{q\rightarrow h}(z_h,\mu^2) \nonumber \\
\Delta D_{q\rightarrow h}(z_h,\mu^2)
&=&\int_0^{\mu^2} \frac{d\ell_T^2}{\ell_T^2} 
\frac{\alpha_s}{2\pi} \int_{z_h}^1 \frac{dz}{z}
\left[ \Delta\gamma_{q\rightarrow qg}(z) 
D_{q\rightarrow h}(\frac{z_h}{z})
+\Delta\gamma_{q\rightarrow gq}(z)
D_{g\rightarrow h}(\frac{z_h}{z})\right] \, , \label{eq:dmod}
\end{eqnarray}
where $D_{a\rightarrow h}(z_h,\mu^2)$ is the leading-twist parton
fragmentation function and
\begin{eqnarray}
\Delta\gamma_{q\rightarrow qg}(z)&=&
\left[\frac{1+z^2}{(1-z)_+}T^A_{qg}(x,x_L) + 
\delta(1-z)\Delta T^A_{qg}(x,\ell_T^2) \right]
\frac{C_A2\pi\alpha_s}
{(\ell_T^2+\langle k_T^2\rangle) N_c f_q^A(x,\mu_I^2)}
\label{eq:dsplit1}\\
\Delta\gamma_{q\rightarrow gq}(z) 
&=& \Delta\gamma_{q\rightarrow qg}(1-z) \label{eq:dsplit2}.
\end{eqnarray}
are the modified splitting functions for the induced gluon radiation. Similar
as the vacuum case, the $\delta$-function part is from the virtual 
correction contribution with $\Delta T^A_{qg}(x,\ell_T^2)$ defined as
\begin{equation}
\Delta T^A_{qg}(x,\ell_T^2) \equiv
\int_0^1 dz\frac{1}{1-z}\left[ 2 T^A_{qg}(x,x_L)|_{z=1}
-(1+z^2) T^A_{qg}(x,x_L)\right] \, . \label{eq:vsplit}
\end{equation}
Such virtual corrections are important to ensure the infrared safety
of the modified fragmentation function and the unitarity of the
gluon radiation processes. This virtual correction is equivalent in
nature to the absorption processes in some effective models \cite{kop}.

To evaluate the modified fragmentation, one needs to know the
two-parton correlation function $T^A_{qg}(x,x_L)$ which consists
of both diagonal and off-diagonal twist-four parton matrices. We
generalize the factorization assumption of LQS \cite{lqs} to both
types of matrices. Assuming a Gaussian nuclear distribution in the 
rest frame,
$\rho(r)\sim \exp(-r^2/2R_A^2)$, $R_A=1.12 A^{1/3}$ fm, 
we express $T_{qg}^A$ in terms of single parton distributions,
\begin{equation}
T^A_{qg}(x,x_L)=\widetilde{C}m_NR_A f_q^A(x) (1-e^{-x_L^2/x_A^2}),
\label{eq:tqg2}
\end{equation}
where $x_A=1/m_NR_A$, and $m_N$ is the nucleon's mass. The off-diagonal 
terms involve transferring momentum $x_L$ between different nucleons 
inside a nucleus and thus should be suppressed for large nuclear size 
or large momentum fraction $x_L$. Notice that $\tau_f=1/x_Lp^+$ is 
the gluon's formation time. Thus, $x_L/x_A=L_A/\tau_f$ with $L_A=R_Am_N/p^+$ 
being the nuclear size in the infinite momentum frame.

Because of the LPM interference effect, the above effective parton 
correlation and the induced gluon emission vanishes when $x_L/x_A\ll 1$.
Therefore, the formation time of the gluon radiation due to the 
LPM interference requires the radiated gluon to have a minimum transverse 
momentum $\ell_T^2\sim Q^2/MR_A\sim Q^2/A^{1/3}$. 
The nuclear corrections to the fragmentation function due to double 
parton scattering will then be in the order of 
$\alpha_s A^{1/3}/\ell_T^2 \sim \alpha_s A^{2/3}/Q^2$, which depends
quadratically on the nuclear size. For large values of
$A$ and $Q^2$, these corrections are leading and yet the requirement
$\ell_T^2\ll Q^2$ for the logarithmic approximation in deriving the 
modified fragmentation function is still valid.

\begin{figure}[htb]
\begin{minipage}[t]{80mm}
\includegraphics[scale=0.43]{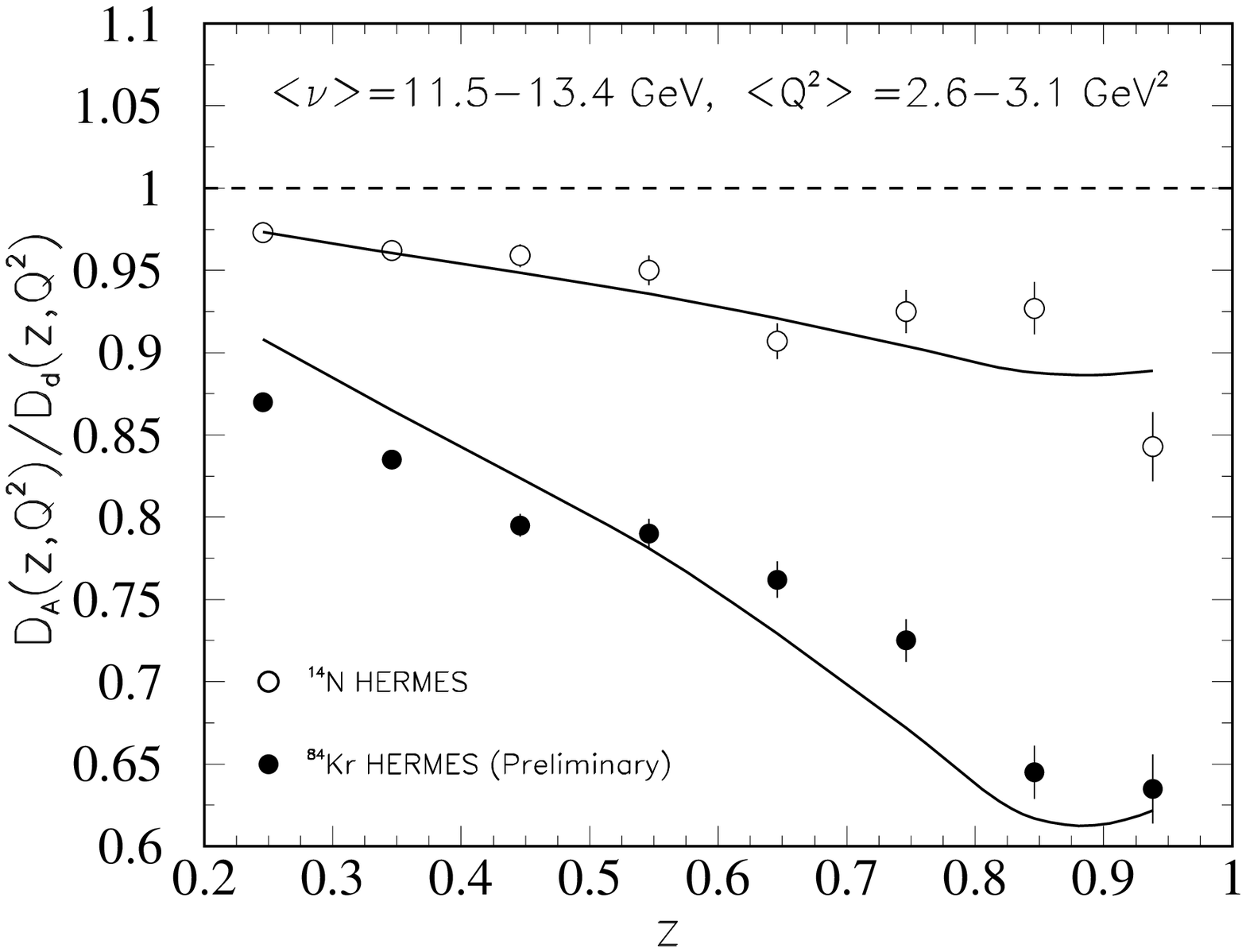}
%\framebox[79mm]{\rule[-26mm]{0mm}{52mm}}
\caption{The predicted nuclear modification of jet fragmentation function
is compared to the HERMES data \protect\cite{hermes}.}
\label{fig3}
\end{minipage}
\hspace{\fill}
\begin{minipage}[t]{75mm}
\includegraphics[scale=0.43]{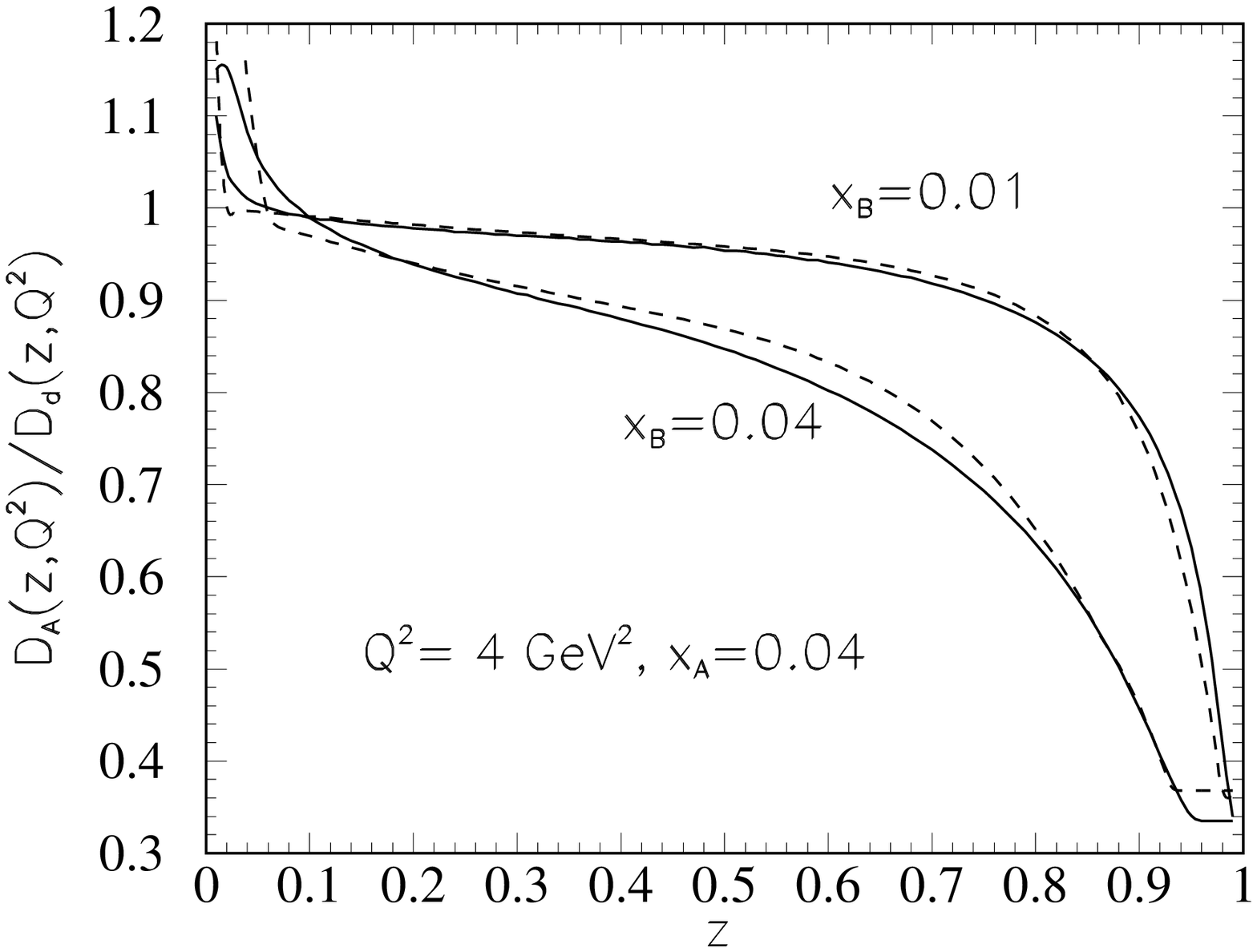}
\caption{Comparison of the calculated nuclear modification with the
effective model in Eq.~(\ref{dbar0}) with $\Delta z=0.6 \langle z_g\rangle$.
}
\label{fig4}
\end{minipage}
\end{figure}

Shown in Fig.~3 are the calculated nuclear modification factor of the
fragmentation function for $^{14}N$ and $^{84}Kr$ targets as compared to the 
recent HERMES data \cite{hermes}. The predicted shape of the $z$ dependence 
and the quadratic nuclear size dependence agrees well with the 
experimental data. The energy 
dependence of the suppression also has excellent agreement with our 
prediction \cite{wang02}. What is amazing is the clear 
quadratic $A^{2/3}$ nuclear size dependence of the suppression
which is a true QCD non-Abelian effect. 
In fitting the data of the overall 
suppression for $^{14}N$ target we obtain the only parameter in 
our calculation, $\widetilde{C}\alpha^2_{\rm s}=0.00065$ GeV$^2$.
This parameter is shown to be related to the transverse momentum
nuclear broadening of Drell-Yan dilepton in $pA$ collisions \cite{gq},
$\langle\Delta q_\perp^2\rangle=\widetilde{C}\pi\alpha_{\rm s}/N_cx_A$.
With an experimental value \cite{gq} of 
$\langle\Delta q_\perp^2\rangle=0.016A^{1/3}$ GeV$^2$ and
with $\alpha_{\rm s}=0.21$ (at $M^2=40$ GeV$^2$), one finds
$\widetilde{C}\alpha^2_{\rm s}=0.00057$ GeV$^2$. This indicates
that $\widetilde{C}$ might have some $Q^2$ dependence.
With this value of $\widetilde{C}$
we can also predict the nuclear transverse momentum broadening of 
single jets in DIS \cite{wang02}.

With the modified fragmentation function in Eq.~(\ref{eq:dmod}), one
can calculate theoretically the average energy loss by the quark,
which is the energy carried away by the radiated gluons,
 \begin{eqnarray}
\Delta E=\nu\langle\Delta z_g\rangle
\approx  \widetilde{C}\alpha_{\rm s}^2 m_NR_A^2 (C_A/N_c) 3\ln(1/2x_B).
\end{eqnarray}
With the value of $\alpha_{\rm s}^2\widetilde{C}$, and $L_A=R_A\sqrt{2\pi}$
one gets the quark energy loss $dE/dx\approx 0.5$ GeV/fm for a $Au$ 
nucleartarget.

\section{JET QUENCHING IN HEAVY-ION COLLISIONS}

In high-energy heavy-ion collisions, the jet production rate is
not affected by the formation of dense matter and the final state
multiple scattering. One can assume that the high $p_T$ hadron
spectra can then be given by the convolution of the jet production
cross section and the medium modified jet fragmentation 
function $\widetilde{D}_{h/c}(z_c,Q^2)$,
\begin{eqnarray}
  \frac{d\sigma^h_{AB}}{dyd^2p_T}&=&K\int d^2b d^2r
  t_A(r)t_B(|{\bf b}-{\bf r}|)\sum_{abcd}
  \int dx_a dx_b d^2k_{aT} d^2k_{bT} 
  g_B(k_{bT},Q^2,|{\bf b}-{\bf r}|) 
  \nonumber \\ &\times & g_A(k_{aT},Q^2,r)
f_{a/A}(x_a,Q^2,r)f_{b/B}(x_b,Q^2,|{\bf b}-{\bf r}|) 
  \frac{\widetilde{D}_{h/c}(z_c,Q^2)}{\pi z_c}
  \frac{d\sigma_{ab\rightarrow cd}}{d\hat{t}}, \label{eq:nch_AA}
\end{eqnarray}
where $t_A(r)$ is the thickness function of the nucleus $A$,
$f_{a/A}(x_a,Q^2,r)$ is the parton distribution in a nucleus,
$g_A(k_{aT},Q^2,r)$ is the distribution of parton intrinsic transverse
momentum with nuclear broadening \cite{wang98}.

In principle, one should use the modified fragmentation function evaluated 
according to the pQCQ calculation for a dense medium. However, before
that can be done in a practical manner, we have used the effective
approach in Eq.~(\ref{dbar0}) by rescaling the fractional momentum by
$1-\Delta z$ to take into account of the parton energy loss. To verify
whether such an effective approach is adequate, we compare the
two modified fragmentation functions in Fig.~\ref{fig4}. We found
that the effective model (dashed lines) can reproduce the pQCD
result (solid lines) of Eq.~(\ref{eq:dmod}) very well, but only when
$\Delta z$ is set to be $\Delta z\approx 0.6 \langle z_g\rangle$.
Therefore the actual averaged parton energy loss should be about
1.6 times of that used in the effective modified fragmentation function.
This difference is caused by the absorptive processes or unitarity
correction effect in the full pQCD calculation.

Unlike in DIS nuclear scattering, the dense medium in high-energy
heavy-ion collisions is not static. It has to go through rapid
expansion which should also affect the effective total parton energy
loss. The total energy loss extracted from experiments should be
a quantity that is averaged over the whole evolution history of the
expanding system. It is therefore useful to convert the averaged quantity
to an energy loss in a static system that has the same parton
density as the expanding system at its initial stage. If the
averaged total parton energy loss in a longitudinally expanding system 
with a transverse size $R$ is $\Delta E_{1{\rm d}}$, one finds \cite{gvw}
that the corresponding parton energy loss in a static system with the
same initial parton density would be 
$\Delta E=\Delta E_{1{\rm d}}(R/2\tau_0)$. Here $\tau_0$ is the initial
formation time of the dense medium.

\begin{figure}[htb]
\begin{minipage}[t]{80mm}
\includegraphics[scale=0.55]{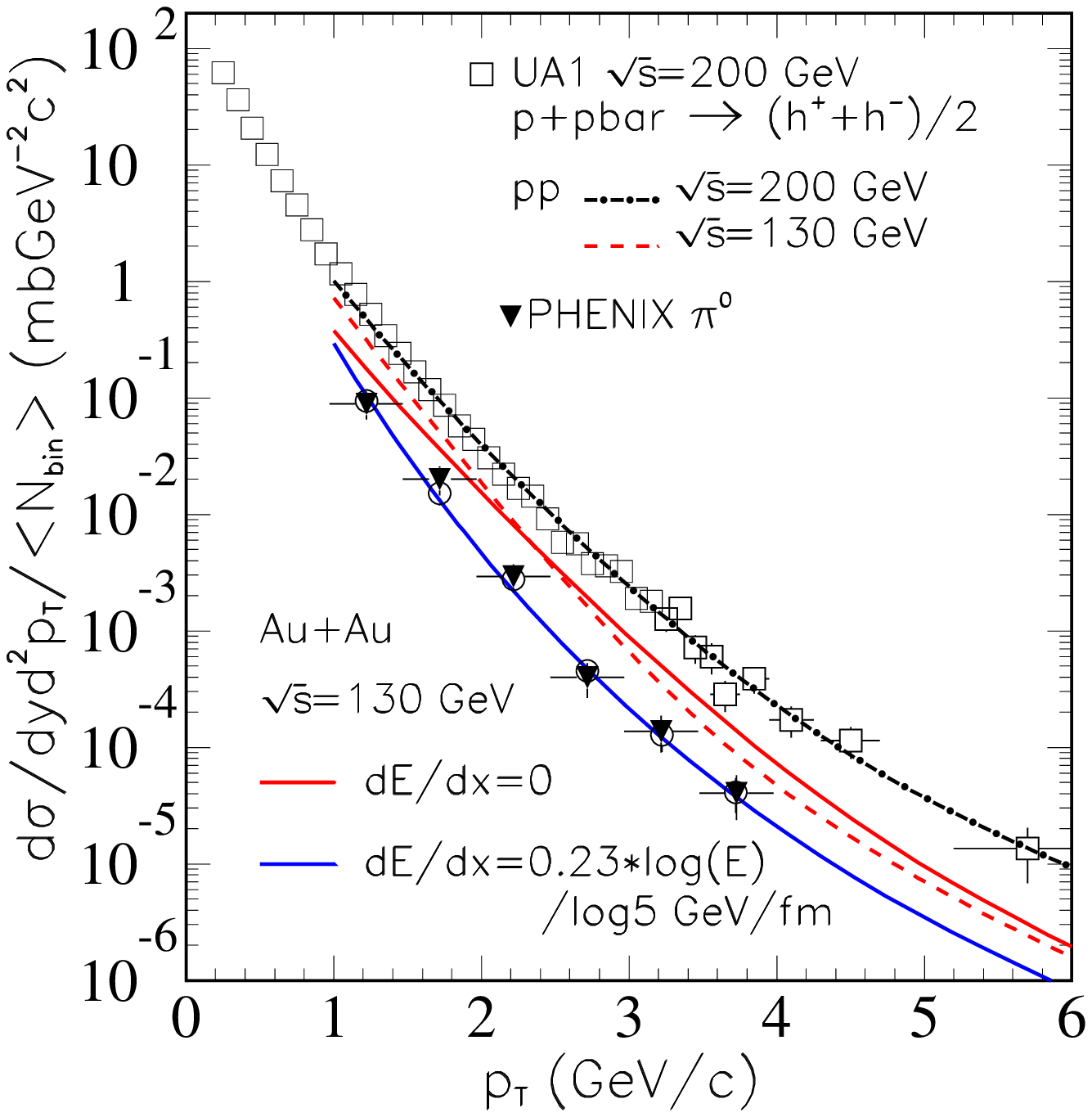}
%\framebox[79mm]{\rule[-26mm]{0mm}{52mm}}
\caption{pQCD parton model calculation of the charged hadron and 
pion spectra in $p\bar{p}$ and central $Au+Au$ collisions compared
with the experimental data \protect\cite{phenix,ua1}. The effective
modified fragmentation function is used in the calculation.}
\label{fig5}
\end{minipage}
\hspace{\fill}
\begin{minipage}[t]{75mm}
\includegraphics[scale=0.43]{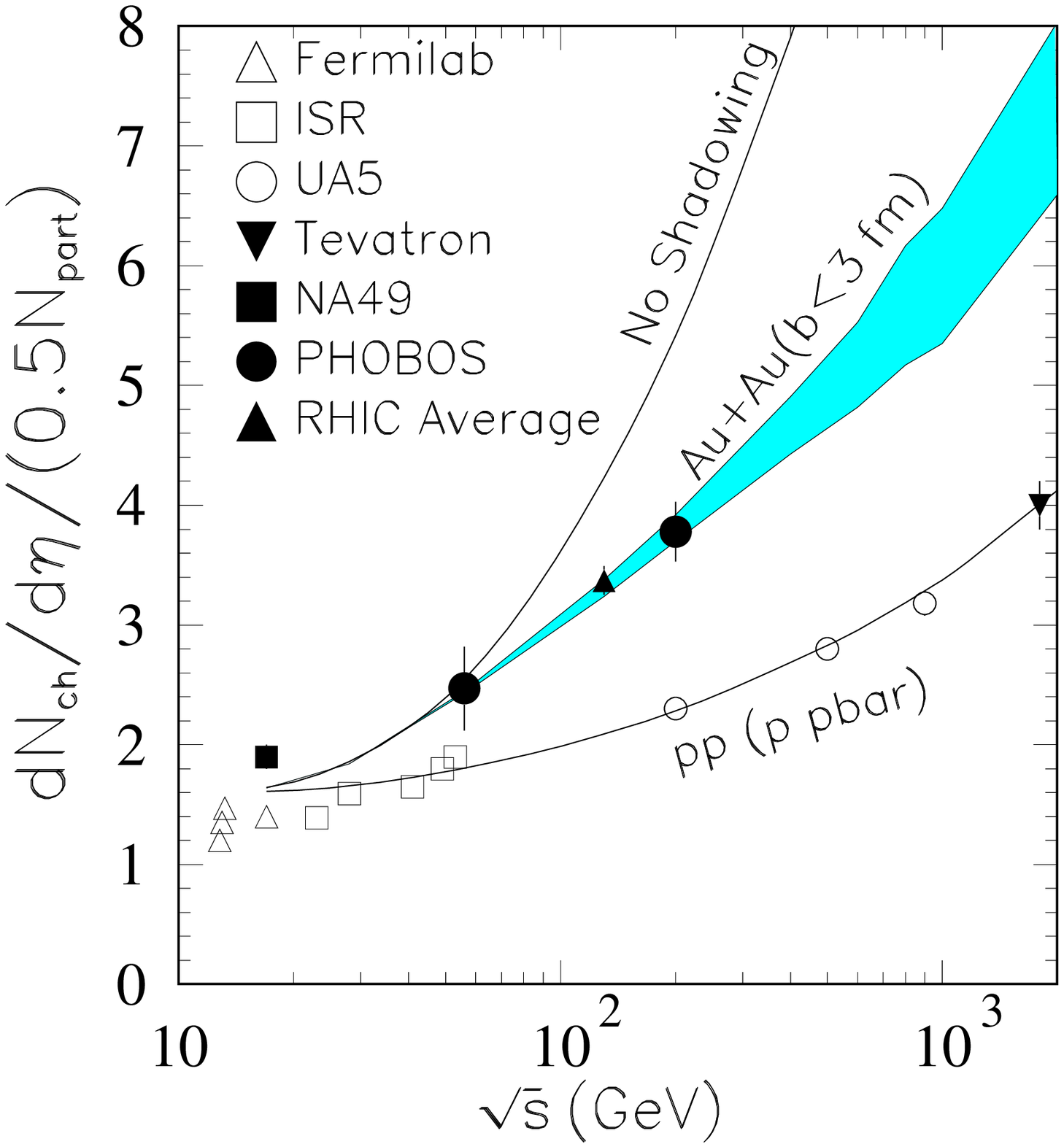}
\caption{The energy dependence of the charged multiplicity
density per participant nucleon pair in the two-component model
with and without parton shadowing compared with the experimental
data \protect\cite{phobos,ppdata}.}
\label{fig6}
\end{minipage}
\end{figure}

Comparing the recent PHENIX data \cite{phenix} with Eq.~(\ref{eq:nch_AA})
as shown in Fig.~\ref{fig5}, one can extract a value of 
$dE/dx=0.23\ln E/\ln 5$GeV/fm with a logarithmic energy dependence
that one needs to use in the effective modified fragmentation function
in fitting the data. Taking into account the unitarity 
correction effect and the expansion, this
corresponds to an effective energy loss 
$dE/dx=1.6\times 0.23 (R/2\tau_0)\ln E/\ln 5$.
in a static system with a density similar to the initial stage of the
expanding system at $\tau_0$. With $R\sim 6$ fm and $\tau_0\sim 1/p_0=0.2$ fm,
this would give $dE/dx\approx 7.3$ GeV/fm for a 10 GeV parton, 
which is about 15 times of
that in a cold nuclear matter as extracted from the DIS processes. Since
the parton energy loss is directly proportional to gluon density, this
implies that the gluon density in the initial stage of $Au+Au$
collisions is about 15 times higher than that inside a cold nucleus.

\section{GLOBAL CONSTRAINTS OF PARTON ENERGY LOSS}

When a fast parton experiences multiple scattering and induced gluon
radiation, the radiated gluon would also contribute to the final
parton production leading to the enhancement of modified fragmentation
function at small $z$ as shown in Fig.~\ref{fig4}. These partons might
get thermalized in the medium, but will contribute to the final hadron
production. In the HIJING model \cite{hijing} with the default setting, 
a simple $dE/dx$=2 GeV/fm is assumed for every jet with $p_T>2 $ GeV/$c^2$.
Such a default setting predicted an enhancement of the total hadron
multiplicity in the central region \cite{wg92} and a strong energy
dependence \cite{wg00} which is now excluded by the new PHOBOS 
data \cite{phobos} of $Au+Au$ collisions at $\sqrt{s}=200$ GeV. This
either implies that the  $dE/dx=2$ GeV/fm used in HIJING is too large
or the onset of jet quenching is set at too low $p_T$ or both. 
According to the
recent study of parton energy loss with detailed balance \cite{ww},
gluon absorption by the fast parton in the thermal medium reduces
the parton energy loss significantly for low energy partons. Taking
into account this strong energy dependence of $dE/dx$, one might
expect that the threshold for gluon radiation could be higher than
what HIJING used. In fact, if one increases the threshold to 3 GeV
in HIJING, one can still fit both $dN_{\rm ch}/d\eta$ and the suppressed
hadron spectra at high $p_T$. Since the production rate of mini-jets
with $p_T>3$ GeV/$c^2$ is very small, their contribution due to
induced bremsstrahlung to the total charged multiplicity is negligible
at the RHIC energies. By similar argument, the effect of parton
thermalization on the total hadron multiplicity might also be small.

Neglecting the effect of jet quenching on the total multiplicity, one
can assume the final $dN{\rm ch}/d\eta$ to be proportional to the total
number of mini-jets produced in addition to the soft particle production.
Assuming that the mini-jet production is proportional to the
number of binary collisions and the soft part is proportional to the
number of participating nucleons, one get in this simple two-component
model
\begin{equation}
\frac{dN_{ch}}{d\eta}=\frac{1}{2}\langle N_{\rm part}\rangle 
\langle n\rangle_{s} 
+ \langle n\rangle_{h}\langle N_{\rm binary}\rangle 
\frac{\sigma_{\rm jet}^{AA}(s)}{\sigma_{\rm in}},
\label{eq:nch} 
\end{equation}
where $\sigma_{\rm jet}^{AA}(s)$ is the averaged inclusive jet cross
section per $NN$ in $AA$ collisions. The energy dependence of
the above multiplicity is shown in Fig.~\ref{fig6}. The parameters
$\langle n\rangle_{\rm s}=1.6$ and $\langle n\rangle_{\rm h}=2.2$
is fixed by the $p+p(\bar{p})$ data. The average number of
participant nucleons and number of binary collisions for given
impact-parameters can be estimated using HIJING Monte Carlo simulation.
If one assumes that the jet production 
cross section $\sigma_{\rm jet}^{AA}(s)$ is the same 
as in $p+p$ collisions, the resultant energy dependence of the 
multiplicity density in central nuclear collisions is much stronger 
than the RHIC data. Therefore, one has to consider nuclear effects 
of jet production in heavy-ion collisions. Using a more recent
parameterization \cite{lw}, we found that the RHIC data requires
a stronger nuclear shadowing of gluon distributions than quarks.

\section{CONCLUSIONS}

We have calculated the medium modification of the jet fragmentation
functions due to gluon radiation induced by the multiple parton
scattering. The predictions of the shape, energy dependence,
and the quadratic nuclear size $A^{2/3}$ dependence of
the modification agree well with the recent HERMES data. The
resultant parton energy loss in the cold nuclear medium is estimated
to be about 0.5 GeV/fm inside $Kr$ nuclei. Comparing to the
QCD result of the modification of fragmentation function, we found
that the actual averaged energy loss is about 1.6 times that of the 
effective energy loss used in a earlier effective model for the same
modification. Considering the effect of expansion, we found that the
recent PHENIX data imply a medium induced energy loss in central $Au+Au$
collisions equivalent to 7.3 GeV/fm in a static medium with the same
gluon density as in the initial stage of the collision. This is about 15
times larger than the energy loss in a cold nucleus. Due to detailed
balance in induced radiation and absorption, we argue that parton
energy loss for small and medium high energy jets is very small. 
Therefore, the contribution from their induced radiation to the 
hadron multiplicity is negligible.

This work is supported by  the Director, Office of Energy 
Research, Office of High Energy and Nuclear Physics, 
Division of Nuclear Physics, and by the Office of Basic Energy Science, 
Division of Nuclear Science, of  the U.S. Department of Energy 
under Contract No. DE-AC03-76SF00098 and in part by 
NSFC under project 19928511.

\end{document}